
\magnification\magstep1

\openup 3\jot

\input mssymb
\def\hbar{\mathchar '26\mkern -9muh}

\catcode`@=11
\def\eqaltxt#1{\displ@y \tabskip 0pt
  \halign to\displaywidth {%
    \rlap{$##$}\tabskip\centering
    &\hfil$\@lign\displaystyle{##}$\tabskip\z@skip
    &$\@lign\displaystyle{{}##}$\hfil\tabskip\centering
    &\llap{$\@lign##$}\tabskip\z@skip\crcr
    #1\crcr}}
\def\eqallft#1{\displ@y \tabskip 0pt
  \halign to\displaywidth {%
    $\@lign\displaystyle {##}$\tabskip\z@skip
    &$\@lign\displaystyle{{}##}$\hfil\crcr
    #1\crcr}}
\catcode`@=12 

\def\pmb#1{\setbox0=\hbox{#1}  \kern-.025em\copy0\kern-\wd0
  \kern0.05em\copy0\kern-\wd0  \kern-.025em\raise.0433em\box0 }
\def\pmbh#1{\setbox0=\hbox{#1} \kern-.12em\copy0\kern-\wd0
            \kern.12em\copy0\kern-\wd0\box0}
\def\sqr#1#2{{\vcenter{\vbox{\hrule height.#2pt
      \hbox{\vrule width.#2pt height#1pt \kern#1pt
         \vrule width.#2pt}
      \hrule height.#2pt}}}}

\def\rchi{{\raise 2pt \hbox {$\chi$}}}
\def\rga{{\raise 2pt \hbox {$\gamma$}}}
\def\rg{{\raise 2 pt \hbox {$g$}}}
\def\({\left(}
\def\){\right)}
\def\<{\left\langle}
\def\>{\right\rangle}

\def\[{\left[}
\def\]{\right]}
\let\text=\hbox
\def\pt{\partial}
\def\eps{\epsilon}
\def\kap{\kappa}

\def\om{\omega}
\def\ol{\overline}

\def\de{\delta}
\def\lam{\lambda}

\def\sig{\sigma}

\def\cH{{\cal H}}

\def\Lam{\Lambda}

\def\wti{\widetilde}

\hfuzz 6pt

\catcode`@=12 
\rightline {\bf DAMTP R93/35}
\centerline {\bf Quantization of the Bianchi type-IX model in}
\centerline {\bf supergravity with a cosmological constant}
\vskip .5 true in
\centerline {A.D.Y. Cheng, P.D. D'Eath and P.R.L.V. Moniz$^*$}
\vskip .5 true in
\centerline {Department of Applied Mathematics and Theoretical Physics}
\centerline {University of Cambridge}
\centerline {Silver Street}
\centerline {Cambridge CB3 9EW}
\centerline {United Kingdom}
\vskip 1 true in
\centerline {\bf ABSTRACT}
\vskip 12 pt
Diagonal Bianchi type-IX models are studied in the quantum theory of $ N = 1 $
supergravity with a cosmological constant. It is shown, by imposing the
supersymmetry and Lorentz quantum constraints, that there are no physical
quantum states in this model. The $ k = + 1 $ Friedmann model in supergravity
with cosmological constant does admit quantum states. However, the Bianchi
type-IX model provides a better guide to the behaviour of a generic state,
since more gravitino modes are available to be excited. These results
indicate that there may be no physical quantum states in the full theory of
$ N = 1 $ supergravity with a non-zero cosmological constant.
\vfill
\centerline {PACs numbers: 04.60.+ $n$, 04.65.+ $e$, 98.80. $Hw$ }
\vskip 6 pt
\noindent
$^*$ e-mail address: prlvm10@amtp.cam.ac.uk
\eject
\noindent
{\bf I. Introduction}

Recently a number of quantum cosmological models have been studied in which
the action is that of supergravity, with possible additional coupling to
supermatter [1--11]. It is sufficient, in finding a physical state, to solve
the Lorentz and supersymmetry constraints of the theory [12,13]. Because of
the anti-commutation relations $ \[ S_A,~\wti S_{A'} \]_+ \sim \cH_{A A'} $,
the supersymmetry constraints $ S_A \Psi = 0,~\ol S_{A'} \Psi = 0 $ on a
physical wave function $ \Psi $ imply the Hamiltonian constraint $ \cH_{A A'}
\Psi = 0 $ [12,13].

In the case of the Bianchi-I model in $ N = 1 $ supergravity with
cosmological constant $ \Lam = 0 $ [8], only two
quantum states appear. Using the factor ordering of [8], one state is $
h^{{1 \over 4}} $ in the bosonic sector, where $ h = \det h_{i j} $ is the
determinant of the three-metric, and the other state is $ h^{- {1 \over 4}} $
in the sector filled with fermions. In the case of Bianchi IX with $ \Lam = 0
$, there are
again two states, of the form $ \exp ( \pm I / \hbar ) $ where $ I $ is a
certain Euclidean action, one in the empty and one in the filled fermionic
sector [9,14].
When the usual choice of spinors constant in the standard basis is made
for the gravitino field, the bosonic state $ \exp ( - I / \hbar ) $ is the
wormhole state [9,15]. With a different choice, one obtains the
Hartle--Hawking state [14,16]. Similar states were found for $ N = 1 $
supergravity in the more general Bianchi models of class $ A $ [10].
[Supersymmetry (as well as other considerations) forbids mini-superspace
models of class $ B $.] It was also found
in the general theory of quantized $ N = 1 $ supergravity with $ \Lam = 0 $
that there are two
bosonic states of the form $ \exp (- I / \hbar) $, where $ I $ is the
wormhole or the Hartle--Hawking classical action [17]. [There may be many
other
bosonic states.] There are also two states of the form $ \exp (I / \hbar) $
in the filled sector.

It is of interest to extend these results, by studying more general locally
supersymmetric actions, initially in Bianchi models. Possibly the simplest
such generalization is the addition of a cosmological constant in $ N = 1 $
supergravity [18]. It was found that in the Bianchi-I case there are no
physical states for $ N = 1 $ supergravity with a $ \Lam $-term [11]. The
Bianchi-IX model with $ \Lam $-term and with $ N = 1 $ supersymmetry in one
dimension was studied by Graham [4]. Here we treat the Bianchi-IX model with
$ \Lam $-term with the full $ N = 4 $ supersymmetry in one dimension. We
shall see that there are again no physical quantum states. The
calculations are described in Sec.~II. We also treat briefly in Sec.~III the
spherical $ k = + 1 $ Friedmann model, and find that there is a two-parameter
family of solutions of the quantum constraints with a $ \Lam $-term.
Nevertheless, as will be seen, the Bianchi-IX model provides a better guide
to the generic result, since more spin-$ {3 \over 2} $ modes are available to
be excited in the Bianchi-IX model, while the form of the fermionic fields
needed for supersymmetry in the $ k =
+ 1 $ Friedmann model is very restrictive [6]. Sec.~IV contains the
Conclusion.
\medbreak
\noindent
{\bf II. QUANTUM STATES FOR THE BIANCHI-IX MODEL WITH A $ \Lam $-TERM}

Using two-component spinors [6,13], the action [18] is
$$ S = \int d^4 x \[ \eqalign {
{}~& \( 2 \kap^2 \)^{- 1} \( \det e \)~\( R - 3 g^2 \) \cr
+ &{1 \over 2} \eps^{\mu \nu \rho \sig} \( \ol \psi^{A'}_{~~\mu} e_{A A' \nu}
D_\rho \psi^A_{~~\sig} + H.c. \) \cr
- &{1 \over 2} g (\det e)~\( \psi^A_{~~\mu} e_{A B'}^{~~~~\mu} e_B^{~~B' \nu}
\psi^B_{~~\nu} + H.c. \) \cr }
\]~. \eqno (2.1) $$
Here the tetrad is $ e^a_{~\mu} $ or equivalently $ e^{A A'}_{~~~~\mu} $. The
gravitino field $ \(\psi^A_{~~\mu}, \ol \psi^{A'}_{~~\mu} \) $ is
an odd (anti-commuting) Grassmann quantity. The scalar
curvature $ R $ and the covariant derivative $ D_\rho $ include torsion.
We define $ \kap^2 =
8 \pi $. Here $ g $ is a constant, and the cosmological constant is $ \Lam =
{3 \over 2} g^2 $.

There are two possible approaches to the quantization of this model. One
possibility is to substitute the Bianchi-IX Ansatz for the geometry $ e^{A
A'}_{~~~~\mu} $ and gravitino field $ \( \psi^A_{~~\mu}, \ol
\psi^{A'}_{~~\mu} \) $ into the action (2.1). The components $ \psi^A_{~~\mu}
e^{B B' \mu} $ and $ \ol \psi^{A'}_{~~\mu} e^{B B' \mu} $ are required to be
spatially constant with respect to the standard triad [19] on the Bianchi-IX
three-sphere. One finds that, in order for the form of the Ansatz to be left
invariant by one-dimensional local supersymmetry transformations, possibly
corrected by coordinate and Lorentz transformations [6], one must study the
general non-diagonal Bianchi-IX model [19]. The reduced action could then be
computed, leading to the Hamiltonian and supersymmetry constraints. Finally
the supersymmetry constraints could be imposed on physical wave functions.
They would be complicated because of the number of parameters needed to
describe the off-diagonal model.

The other alternative, taken here, is to apply the supersymmetry constraints
of the general theory at a diagonal Bianchi-IX geometry [9]. This is valid
since the supersymmetry constraints are of first order in bosonic
derivatives, and give expressions such as $ \de \Psi / \de h_{i m} (x) $ in
terms of known quantities and $ \Psi $. These equations can be evaluated at a
diagonal Bianchi-IX geometry, parametrized by three radii $ A,~B,~C $. One
multiplies (e.g.) by $ \de h_{i m} (x) = \pt h_{i m} / \pt A $ and integrates $
\int d^3 x (~~) $ to obtain an equation for $ \pt \Psi / \pt A $ in terms of
known
quantities. The need to consider off-diagonal metrics is thereby avoided.

The general classical supersymmetry constraints are, with the help of [13],
seen to be
$$ \ol S_{A'} = g h^{1 \over 2} e^{A~~~i}_{~~A'} n_{A B'} \ol \psi^{B'}_{~~i}
+ \eps^{i j k} e_{A A' i}~^{3 s} D_j \psi^A_{~~k} + {1 \over 2} i \kap^2
\psi^A_{~~i} p_{A A'}^{~~~~i}~, \eqno (2.2) $$
and the conjugate $ S_A $. Here $ n^{A A'} $ is the spinor version of the
unit future-pointing normal $ n^\mu $ to the surface $ t = {\rm const} $. It
is a function of the $ e^{A A'}_{~~~~i} $, defined by
$$ n^{A A'} e_{A A' i} = 0~, \ \ \ \ n^{A A'} n_{A A'} = 1~. \eqno (2.3) $$
In Eq.~(2.2), $ p_{A A'}^{~~~~i} $ is the momentum conjugate to $ e^{A
A'}_{~~~~i} $. The expression $ ^{3 s}D_j $ denotes the three-dimensional
covariant derivative without torsion. Since the components of $ \psi^A_{~~k}
$ are taken to be constant in the Bianchi-IX basis, one can replace $
^{3s}D_j \psi^A_{~~k} $ by $ \om^A_{~~B j} \psi^B_{~~k} $, where $ \om^A_{~~B
j} $ gives the torsion-free connection [13].

The corresponding quantum constraints read, with the help of [13],
$$ \eqalignno {
\ol S_{A} \Psi &= - i \hbar g h^{1 \over 2} e^{A~~~i}_{~~A'} n_{A B'} D^{B
B'}_{~~~~j i} \( h^{1 \over 2} {\pt \Psi \over \pt \psi^B_{~~j}} \) \cr
{}~&+ \eps^{i j  k} e_{A A' i} \om^A_{~~B j} \psi^B_{~~k} \Psi - {1 \over 2}
\hbar
\kap^2 \psi^A_{~~i} {\de \Psi \over \de e^{A A'}_{~~~~i}} = 0~, &(2.4) \cr
S_A \Psi &= g h^{1 \over 2} e_A^{~~A' i} n_{B A'} \psi^B_{~~i} \Psi - i \hbar
\om_{A~~i}^{~~B} \( h^{1 \over 2} {\pt \Psi \over \pt \psi^B_{~~i}} \) \cr
{}~&+ {1 \over 2} i \hbar^2 \kap^2 D^{B A'}_{~~~~j i} \( h^{1 \over 2} {\pt
\over
\pt \psi^B_{~~j}} \)~{\de \Psi \over \de e^{A A'}_{~~~~i}} = 0~. &(2.5) \cr
} $$
Here
$$ D^{B A'}_{~~~~j i} = - 2 i h^{- {1 \over 2}} e^{B B'}_{~~~~i} e_{C B' j}
n^{C A'}~, \eqno (2.6) $$
and $ \pt / \pt \psi^B_{~~j} $ denotes the left derivative [20]. We have made
the replacement $ \de \Psi / \de \psi^B_{~~j} \longrightarrow h^{1 \over 2}
\pt \Psi / \pt \psi^B_{~~j} $. The $ h^{1 \over 2} $ factor ensures that each
term has the correct weight in the equations. One can also check that this
replacement gives the correct supersymmetry constraints in the $ k = + 1 $
Friedmann model (without $ \Lam $-term), where the model was quantized using
the alternative approach via a supersymmetric Ansatz [6].

In addition to the supersymmetry constraints, a physical state $ \Psi $ must
also obey the Lorentz constraints
$$ J^{A B} \Psi = 0~, \ \ \ \ \ol J^{A' B'} \Psi = 0~. \eqno (2.7) $$
These imply that $ \Psi $ is formed from the three-metric $ h_{i j} $ and
from scalar invariants in the gravitino field. To specify this, note the
decomposition [11] of $ \psi^A_{~~B B'} = e_{B B'}^{~~~~i} \psi^A_{~~i} $:
$$ \psi_{A B B'} = - 2 n^C_{~~B'} \rga_{A B  C} + {2 \over 3} \( \beta_A
n_{B B'} + \beta_B n_{A B'} \) - 2 \eps_{A B} n^C_{~~B'} \beta_C~, \eqno
(2.8) $$
where $ \rga_{A B C} = \rga_{(A B C)} $ is totally symmetric and $
\eps_{A B} $ is the alternating spinor. The general Lorentz-invariant wave
function is a polynomial of sixth degree in Grassmann variables:
\vfill\eject
$$ \eqalignno {
\Psi \( e^{A A'}_{~~~~i},~\psi^A_{~~i} \) &= \Psi_0 \( h_{i j} \) + \( \beta_A
\beta^A \) \Psi_{21} \( h_{i j} \) + \( \rga_{A B C} \rga^{A B C} \)
\Psi_{22} \( h_{i j} \) \cr
{}~&+ \( \beta_A \beta^A \)~\( \rga_{B C D} \rga^{B C D} \) \Psi_{41} \( h_{i
j} \) + \( \rga_{A B C} \rga^{A B C} \)^2 \Psi_{42} \( h_{i j} \) \cr
{}~&+ \( \beta_A \beta^A \)~\( \rga_{B C D} \rga^{B C D} \)^2 \Psi_6 \( h_{i
j} \)~. &(2.9) \cr
} $$
As described in [11], any other Lorentz-invariant fermionic polynomials can
be written in terms of these.

We now proceed to solve the supersymmetry and Lorentz constraints.
The diagonal Bianchi-IX three-metric is given in terms of the three radii $
A, B, C $ by
$$ h_{i j} = A^2 E^1_{~i} E^1_{~j} + B^2 E^2_{~i} E^2_{~j} + C^2 E^3_{~i}
E^3_{~j}~, \eqno (2.10) $$
where $ E^1_{~i}, E^2_{~i}, E^3_{~i} $ are a basis of unit left-invariant
one-forms on the three-sphere [19]. In the calculation, we shall repeatedly
need the expression, formed from the connection:
$$ \eqalignno {
\om_{A B i} n^A_{~~B'} e^{B B' j} &= {i \over 4} \( {C \over A B} + {B \over C
A} - {A \over B C} \) E^1_{~i} E^{1 j} \cr
{}~&+ {i \over 4} \( {A \over B C} + {C \over A B} - {B \over C A} \) E^2_{~i}
E^{2 j} \cr
{}~&+ {i \over 4} \( {B \over C A} + {A \over B C} - {C \over A B} \) E^3_{~i}
E^{3 j} &(2.11) \cr } $$
This can be derived from the expressions for $ \om^{A B}_{~~~i} $ given in
[9,13].

First consider the $ \ol S_{A'} \Psi = 0 $ constraint at the level $ \psi^1 $
in powers of fermions. One obtains
$$ {3 \over 16} \hbar g h^{1 \over 2} e_{B A'}^{~~~~i} \psi^B_{~~i} \Psi_{21}
+ \eps^{j k i} e_{A A' j} \om^A_{~~B k} \psi^B_{~~i} \Psi_0 + \hbar \kap^2
e_{B A' j} \psi^B_{~~i} {\de \Psi_0 \over \de h_{i j}} = 0~. \eqno (2.12) $$
Since this holds for all $ \psi^B_{~~i} $, one can conclude
$$ {3 \over 16} \hbar g h^{1 \over 2} e_{B A'}^{~~~~i} \Psi_{21} + \eps^{j k
i} e_{A A' j} \om^A_{~~B k} \Psi_0 + \hbar \kap^2 e_{B A' j} {\de \Psi_0
\over \de h_{i j}} = 0~. \eqno (2.13) $$
Now multiply this equation by $ e^{B A' m} $, giving
$$ - {3 \over 16} \hbar g h^{i m} h^{1 \over 2} \Psi_{21}
+ \eps^{j k i} e_{A A' j} e^{B A' m} \om^A_{~~B k} \Psi_0
- \hbar \kap^2 {\de \Psi_0 \over \de h_{i m}} = 0~. \eqno (2.14) $$
The second term can be simplified using [6]
$$ e_{A A' j} e^{B A'}_{~~~~m} = - { 1 \over 2} h_{j m} \eps_A^{~~B} + i
\eps_{j m n}
h^{1 \over 2} n_{A A'} e^{B A' n}~. \eqno (2.15) $$
One then notes, as above, that by taking a variation among the Bianchi-IX
metrics, such as
$$ \de h_{i j} = {\pt h_{i j} \over \pt A} = 2 A E^1_{~i} E^1_{~j}~, \eqno
(2.16) $$
multiplying by $ \de \Psi_0 / \de h_{i j} $ and integrating over the
three-geometry, one obtains $ \pt \Psi_0 / \pt A $. Putting this information
together one obtains the constraint
$$ \hbar \kap^2 {\pt \Psi_0 \over \pt A} + 1 6 \pi^2 A \Psi_0 + 6 \pi^2 \hbar
g B C \Psi_{21} = 0~, \eqno (2.17) $$
and two others given by cyclic permutation of $ A B C $.

Next we consider the $ S_A \Psi = 0 $ constraint at order $ \psi^1 $. One
uses the relations $ \pt \( \beta_A \beta^A \) / \pt \psi^B_{~~i} = -
n_A^{~~B'} e_{B B'}^{~~~~i} \beta^A $ and $ \pt \( \rga_{A D C} \rga^{A D C}
\) / \pt \psi^B_{~~i} = - 2 \rga_{B D C}~ n^{C C'} e^{D~~~~~i}_{~~~C'} $, and
writes out $ \beta^A $ and $ \rga_{B D C} $ in terms of $ e^{E E'}_{~~~~j} $
and $ \psi^E_{~~j} $. Proceeding by analogy with the previous calculation
above, one again `divides out' by $ \psi^B_{~~j} $ to obtain
\vfill\eject
$$ g h^{1 \over 2} e_A^{~~A' j} n_{B A'} \Psi_0
- {1 \over 4} i \hbar \om_{A~~i}^{~~C} h^{1 \over 2} e_{C B'}^{~~~~i}
e_B^{~~B' j} \Psi_{21} $$
$$ - \( \eqalign {
&{1 \over 3} i \hbar \om_{A B i} h^{1 \over 2} e_{D A'}^{~~~~~j} e^{D A' i}
\cr
&+ {2 \over 3} i \hbar \om_{A~~i}^{~~E} h^{1 \over 2} e_{E A'}^{~~~~j}
e_B^{~~A' i} \cr } \) \Psi_{22} $$
$$ + {1 \over 4} \hbar^2 \kap^2 e^C_{~~B' i} n_C^{~~A'} e_B^{~~B' j} e_{A A'
m} {\de \Psi_{21} \over \de h_{i m}} $$
$$ - 2 \hbar^2 \kap^2 \( \eqalign {
&- {2 \over 3} \de_i^{~j} n_B^{~~A'} + {1 \over 3} e_{B~~~i}^{~~C'} n^{C
A'} e_{C C'}^{~~~~j} \cr
&- {1 \over 6} n_C^{~~B'} e^{C A'}_{~~~~i} e_{B B'}^{~~~~j} - {1 \over 6}
n_B^{~~B'} e^{C A'}_{~~~~i} e_{C B'}^{~~~~j} \cr } \)
e_{A A'm} {\de \Psi_{22} \over \de h_{i m}} = 0~. \eqno (2.18) $$
One replaces the free spinor indices $ A B $ by the spatial index $ n $ on
multiplying by $ n^A_{~~D'} e^{B D' n} $, giving
$$ - {1 \over 2} g h^{1 \over 2} h^{j n} \Psi_0 $$
$$ + {1 \over 8} i \hbar h^{1 \over 2} \( \eqalign {
&h^{i j} \om_{A B i} n^A_{~~B'} e^{B B' n} - h^{i n} \om_{A B i} n^A_{~~B'}
e^{B B' j} \cr
&+ h^{j n} \om_{A B i} n^A_{~~B'} e^{B B' i} \cr } \) \Psi_{21} $$
$$ + {1 \over 3} i \hbar h^{1 \over 2} \( \eqalign {
&2 h^{i j} \om_{A B i} n^A_{~~B'} e^{B B' n} + h^{i n} \om_{A B i} n^A_{~~B'}
e^{B B' j} \cr
&- h^{j n} \om_{A B i} n^A_{~~B'} e^{B B' i} \cr } \) \Psi_{22} $$
$$ \eqalignno {
&+ {1 \over 16} \hbar^2 \kap^2 \( \de_i^{~j} \de_m^{~n} - \de_i^{~n}
\de_m^{~j} + h_{i m} h^{j n} \) {\de \Psi_{21} \over \de h_{i m}} \cr
&- {1 \over 3} \hbar^2 \kap^2 \( 2 \de_i^{~j} \de_m^{~n} + \de_i^{~n}
\de_m^{~j} - h_{i m} h^{j n} \) {\de \Psi_{22} \over \de h_{i m}} = 0~.
&(2.19) \cr } $$
Multiplying by different choices $ \de h_{i m} = \pt h_{i m} / \pt A $ etc.
and integrating over the manifold, one finds the constraints
$$ \eqalignno {
&{1 \over 16} \hbar^2 \kap^2 A^{- 1} \( A {\pt \Psi_{21} \over \pt A} + B
{\pt \Psi_{21} \over \pt B} + C {\pt \Psi_{21} \over \pt C} \) \cr
- &{1 \over 3} \hbar \kap^2 \[ 3 {\pt \Psi_{22} \over \pt A} - A^{- 1} \( A
{\pt \Psi_{22} \over \pt A} + B {\pt \Psi_{22} \over \pt B} + C {\pt
\Psi_{22} \over \pt C} \) \] \cr
&- 16 \pi^2 g B C \Psi_0
- \pi^2 \hbar B C \( {A \over B C} + {B \over C A} + {C \over A B} \)
\Psi_{21} \cr
&+ {1 \over 3} \( 16 \pi^2 \) \hbar B C \( {2 A \over B C} - {B \over C A} -
{C \over A B} \) \Psi_{22} = 0~. &(2.20) \cr
} $$
and two more equations given by cyclic permutation of $ A B C $.

Now consider the $ \ol S_{A'} \Psi = 0 $ constraint at order $ \psi^3 $. It
will turn out that we need go no further than this. The constraint can be
written as
$$ \eqalignno {
&{1 \over 2} \hbar g h^{1 \over 2} e^B_{~~A' j} n_C^{~~B'} e_{B B'}^{~~~~j}
\beta_C \( \rga_{D E F} \rga^{D E F} \) \Psi_{41} \cr
&+ \eps^{i j k} e_{A A' i} \om^A_{~~B j} \psi^B_{~~k} \[ \( \beta_C \beta^C
\) \Psi_{21} + \( \rga_{C D E} \rga^{C D E} \) \Psi_{22} \] \cr
&- {1 \over 2} \hbar^2 \kap^2 \psi^A_{~~i} \[ \( \beta_C \beta^C \) {\de
\Psi_{21} \over \de e^{A A'}_{~~~~i}} + \( \rga_{C D E} \rga^{C D E} \) {\de
\Psi_{22} \over \de e^{A A'}_{~~~~i}} \] = 0~. &(2.21) \cr
} $$
The terms $ \psi^B_{~~k} $ and $ \psi^A_{~~i} $ in the last two lines can be
rewritten in terms of $ \beta_A $ and $ \rga_{F G H} $, using Eq.~(2.8). Then
one can set separately to zero the coefficient of $ \beta^C \( \rga_{D E F}
\rga^{D E F}
\) $, the symmetrized coefficient of $ \rga_{D E F} \( \beta_C \beta^C \) $
and the symmetrized coefficient of $ \rga_{F G H} \( \rga_{C D E} \right .
$ \break $ \left . \rga^{C D E} \) $. These three equations give
$$ {3 \over 4} \hbar g h^{1 \over 2} n^C_{~~A'} \Psi_{41}
- {8 \over 3} \eps^{i j k} e_{A A' i} \om^A_{~~B j} n^B_{~~C'} e^{C
C'}_{~~~~k} \Psi_{22}
+ {4 \over 3} \hbar \kap^2 n^A_{~~B'} e^{C B'}_{~~~~i} {\de \Psi_{22} \over
\de e^{A A'}_{~~~~i}} = 0~, \eqno (2.22) $$
$$ 2 \eps^{i j k} e_{A A' i} \om^A_{~~B j} n^D_{~~B'} e^{C B'}_{~~~~k}
\Psi_{21}
- \hbar \kap^2 n^D_{~~B'} e^{C B'}_{~~~~i} {\de \Psi_{21} \over \de e^{B
A'}_{~~~~i}} $$
$$ + \( B C D \to C D B \) + \( B C D \to D B C \) = 0~, \eqno (2.23) $$
and Eq.~(2.23) with $ \Psi_{21} $ replaced by $ \Psi_{22} $. Contracting
Eq.~(2.22) with $ n_C^{~~A'} $ and integrating over the three-surface gives
$$ {3 \over 4} \( 16 \pi^2 \) \hbar g A B C \Psi_{41} + {2 \over 3} \( 16
\pi^2 \)~ \( A^2 + B^2 + C^2 \) \Psi_{22} $$
 $$ + {2 \over 3} \hbar \kap^2 \( A {\pt \Psi_{22} \over \pt A} + B {\pt
\Psi_{22} \over \pt B} + C {\pt \Psi_{22} \over \pt C} \) = 0~. \eqno (2.24)
$$
Contracting Eq.~(2.23) with $ e^{B A' \ell} n_{C C'} e_D^{~~C' N} $,
multiplying by $ \de h_{\ell n} = \pt h_{\ell n} / \pt A $ and integrating
gives
$$ 3 \hbar \kap^2 {\pt \Psi_{21} \over \pt A} - \hbar \kap^2 A^{- 1} \( A
{ \pt \Psi_{21} \over \pt A} + B {\pt \Psi_{21}
\over \pt B} + C {\pt \Psi_{21} \over \pt C} \) $$
$$ - 16 \pi^2 B C \( {C \over A B} + {B \over C A} - 2 {A \over B C} \) = 0~,
\eqno (2.25) $$
and two more equations given by permuting $ A B C $ cyclically. The equation
(2.25) also holds with $ \Psi_{21} $ replaced by $ \Psi_{22} $.

There is a duality between wave functions $ \Psi \( e^{A A'}_{~~~~i},
\psi^A_{~~i} \) $ and wave functions $ \wti \Psi \( e^{A A'}_{~~~~i}
\right . $,
$ \left .
\wti \psi^{A'}_{~~i} \) $, given by a fermionic Fourier transform [13]. The
$ S_A $ and $ \ol S_{A'} $ operators interchange r\^oles under this
transformation, and the r\^oles of $ \Psi_0 $ and $ \Psi_6,~\Psi_{21} $ and $
\Psi_{42} $, and $ \Psi_{22} $ and $ \Psi_{41} $ are interchanged. We shall
proceed by showing that $ \Psi_{22},~\Psi_{21} $ and $ \Psi_0 $ must vanish
for $ g \ne 0 $ (or $ \Lam \ne 0 $), and hence by the duality the entire
wave function must be zero.

Consider first the equation (2.25) and its permutations for $ \Psi_{21} $ and
$ \Psi_{22} $. One can check that these are equivalent to
$$ \hbar \kap^2 \( A {\pt \Psi_{21} \over \pt A} - B {\pt \Psi_{21} \over \pt
B} \) = 16 \pi^2 \( B^2 - A^2 \) \Psi_{21} \eqno (2.26) $$
and cyclic permutations. One can then integrate Eq.~(2.26) along a
characteristic $ A B = $ const., $ C = $ const., using the parametric
description $ A = w_1 e^\tau $, $ B = w_2 e^{- \tau} $, to obtain
$$ \Psi_{21} = h_1 (A B, C) \exp \[ - {8 \pi^2 \over \hbar
\kap^2} ~ \( A^2 + B^2 \) \]~. \eqno (2.27) $$
The general solution of
$$ \hbar \kap^2 \( B {\pt \Psi_{21} \over \pt B} - C {\pt \Psi_{21}
\over \pt C} \) = 16 \pi^2 \( C^2 - B^2 \) \Psi_{21} \eqno (2.28) $$
is similarly
$$ \Psi_{21} = h_2 (B C, A) \exp \[ - {8 \pi^2 \over \hbar
\kap^2} ~\( B^2 + C^2 \) \]~. \eqno (2.29) $$
Eqs.~(2.27) and (2.29) are only consistent if $ \Psi_{21} $ has the form
$$ \Psi_{21} = F (A B C) \exp \[ - {8 \pi^2 \over \hbar
\kap^2} ~\( A^2 + B^2 + C^2 \) \]~. \eqno (2.30) $$
Similarly
$$\Psi_{22} = G (A B C) \exp \[ - {8 \pi^2 \over \hbar
\kap^2} ~\( A^2 + B^2 + C^2 \) \]~. \eqno (2.31) $$

Substituting Eqs.~(2.30),(2.31) into Eq.~(2.20), one obtains
$$ \eqalignno {
&16 \pi^2 g \Psi_0
= - 2 \pi^2 \hbar (A B C)^{- 1} \( A^2 + B^2 + C^2 \)~(\exp) F \cr
&+ {3 \over 16} \hbar^2 \kap^2 (\exp) F'
+ {2 \over 3} \( 16 \pi^2 \) \hbar (A B C)^{- 1} \( 2 A^2 - B^2 - C^2
\)~(\exp) G &(2.32) \cr } $$
and cyclically, where
$$ \exp = \exp \[ - {8 \pi^2 \over \hbar^2 \kap^2}~\( A^2
+ B^2 + C^2 \) \]~. \eqno (2.33) $$
Now $ \Psi_0 $ should be invariant under permutations of $ A, B, C $. Hence $
G = 0 $. I.e.
$$ \Psi_{22} = 0~. \eqno (2.34) $$
The equation (2.32) and its cyclic permutations, with $ \Psi_{22} = 0 $, must
be solved consistently with Eq.~(2.17) and its cyclic permutations.
Eliminating $ \Psi_0 $, one finds
$$ \eqalignno {
&{3 \hbar^3 \kap^4 \over 16 \( 16 \pi^2 g \)} F'' - {\hbar^2 \kap^2 \over 8
g}~{\( A^2 + B^2 + C^2 \) \over A B C} F' \cr
&+ 6 \pi^2 \hbar g F - {\hbar^2 \kap^2 \over 4 g}~{1 \over B^2 C^2} F +
{\hbar^2 \kap^2 \over 8 g}~{\( A^2 + B^2 + C^2 \)
\over (A B C)^2} F = 0~, &(2.35) \cr } $$
and cyclic permutations. Since $ F = F (A B C) $ is invariant under
permutations, the $ (B C)^{- 2} F $ term and its permutations imply $ F = 0 $.
Thus
$$ \Psi_{21} = 0~. \eqno (2.36) $$
Hence, using Eq.~(2.32),
$$ \Psi_0 = 0~. $$

Then we can argue using the duality mentioned earlier, to conclude that
$$ \Psi_{41} = \Psi_{42} = \Psi_6 = 0~. \eqno (2.37) $$
Hence there are no physical quantum states obeying the constraint equations
in the diagonal Bianchi-IX model. This result will be discussed further in the
Conclusion.

This shows that the Chern--Simons semi-classical wave function of Sano and
Shiraishi [21] for $ N = 1 $ supergravity with $ \Lam $-term can only be an
approximate, and not an exact state in the Bianchi-IX model. If it were
exact, then one could make a Fourier transformation from the Ashtekar
variables used in [21] to the variables $ A, B, C $
used here, to find a non-trivial solution.
\bigbreak
\noindent
{\bf III THE $ k = + 1 $ FRIEDMANN MODEL WITH  $ \Lam $-TERM}

The $ k = + 1 $ Friedmann model without a $ \Lam $ term has been discussed
in [2,6]. There are two linearly independent physical quantum states. One is
bosonic and corresponds to the wormhole state [15], the other is at quadratic
order in fermions and corresponds to the Hartle--Hawking state [16]. In the
Friedmann model with $ \Lam $ term, the coupling between the different
fermionic levels `mixes up' this pattern [4].

In the Friedmann model, the wave function has the form [6]
$$ \Psi = \Psi_0 (A) + \( \beta_C \beta^C \) \Psi_2 (A)~. \eqno (3.1) $$
As part of the Ansatz of [6], one requires $ \psi^A_{~~i} = e^{A A'}_{~~~~i}
\wti \psi_{A'} $ and $ \wti \psi^A_{~~i} = e^{A A'}_{~~~~i} \psi_A $; this is
in order that the form of the one-dimensional Ansatz should be preserved
under one-dimensional local supersymmetry, suitably modified by local
coordinate and Lorentz transformations. Thus the gravitino field is truncated
to spin $ {1 \over 2} $. Note that $ \beta^A = {3 \over 4} n^{A A'} \wti
\psi_{A'} $.

One then proceeds as in Sec.~II to derive the consequences of
the $ \ol S_{A'} \Psi = 0 $ and $ S_A \Psi = 0 $ constraints at level $
\psi^1 $, by writing down the general expression for a constraint and then
evaluating it at a Friedmann geometry. Note that it is not equivalent to
set $ A = B = C $ in Eqs.~(2.17) and (2.20); the coefficients in the
constraint equations are different. One then obtains
$$ \hbar \kap^2 {d \Psi_0 \over d A} + 4 8 \pi^2 A \Psi_0 + 1 8 \pi^2 \hbar g
A^2 \Psi_2 = 0 \eqno (3.2) $$
and
$$ \hbar^2 \kap^2 {d \Psi_2 \over d A} - 4 8 \pi^2 \hbar A \Psi_2 - 2 5 6
\pi^2 g A^2 \Psi_0 = 0~. \eqno (3.3) $$
These give second-order equations, for example
$$ A {d^2 \Psi_0 \over d A^2} - 2 {d \Psi_0 \over d A}
+ \[ - {4 8 \pi^2 \over \hbar \kap^2} A - {(48)^2 \pi^4 \over \hbar^2
\kap^4} A^3 + {9 \times 512 \pi^4 g^2 \over \hbar^2 \kap^4} A^5 \] \Psi_0 =
0~. \eqno (3.4) $$
This has a regular singular point at $ A = 0 $, with indices $ \lam = 0 $ and
3. There are two independent solutions, of the form
$$ \eqalignno {
\Psi_0 &= a_0 + a_2 A^2 + a_4 A^4 + \ldots~, \cr
\Psi_0 &= A^3 \( b_0 + b_2 A^2 + b_4 A^4 + \ldots \) ~, &(3.5) \cr } $$
convergent for all $ A $. They obey complicated recurrence relations, where
(e.g.) $ a_6 $ is related to $ a_4,~a_2 $ and $ a_0 $.

One can look for asymptotic solutions of the type $ \Psi_0 \sim \( B_0 +
\hbar B_1 + \hbar^2 B_2 + \ldots \) \exp $ \break
$ ( - I / \hbar ) $, and finds
$$ I = \pm {\pi^2 \over g^2} \( 1 - 2 g^2 A^2 \)^{3 \over 2}~, \eqno (3.6) $$
for $ 2 g^2 A^2 < 1 $.
The minus sign in $ I $ corresponds to taking the action of the classical
Riemannian solution filling in smoothly inside the three-sphere, namely a
portion of the four-sphere $ S^4 $ of constant positive curvature. This gives
the Hartle--Hawking state [16]. For $ A^2 > \( 1 / 2 g^2 \) $, the Riemannian
solution joins onto the Lorentzian solution [22]
$$ \Psi \sim \cos \left \{ \hbar^{- 1} \[ {\pi^2 \( 2 g^2 A^2 - 1 \)^{3 \over
2} \over g^2} - {\pi \over 4} \] \right \}~, \eqno (3.7) $$
which describes de Sitter space-time.
\bigbreak
\noindent
{\bf IV CONCLUSION}

We have seen here that there are no physical quantum states for $ N  = 1 $
supergravity with a $ \Lam $-term, in the diagonal Bianchi-IX model. The same
result was found for non-diagonal Bianchi-I models in [11]. The physical
states found in Sec.~III for the $ k = + 1 $ Friedmann model, where the
degrees of freedom carried by the gravitino field are $ \beta_A $, disappear
when the further fermionic degrees of freedom $ \rga_{A B C} $ of the
Bianchi-IX model are included.

One could also study this from the point of view of perturbation theory about
the $ k = + 1 $ Friedmann model. As well as the usual gravitational harmonics
[23], gravitino harmonics can be used [24]. For example, the Bianchi-IX model
with radii $ A, B, C $ close together describes a particular type of
`gravitational wave' distortion of the Friedmann model; similarly for the $
\rga_{A B C} $ of the Bianchi-IX model, which describes a particular
`gravitino wave' distortion. Quite generally, in perturbation theory [23,25]
one expects to find a wave function which is a product of the background wave
function $ \Psi (A) $ times an infinite product of wave functions $ \psi_n
$
(perturbations) where $ n $ labels the harmonics. And one further expects that
the
perturbation wave function corresponding to the Bianchi-IX modes must be zero,
by a perturbative version of the argument of Sec.~II. [It will be
interesting to investigate this.] Hence the complete perturbative wave
function should be zero; then physical states would be forbidden for a generic
model of the gravitational and gravitino fields with $ \Lam $-term. This
suggests that the full theory of $ N = 1 $ supergravity with a
non-zero $ \Lam $-term should have no physical states.
\bigbreak
\noindent
{\bf ACKNOWLEDGEMENTS}

A.D.Y.C. thanks the Croucher Foundation of Hong Kong for financial support.
\break
P.R.L.V.M. gratefully acknowledges the support of a Human Capital and Mobility
grant from the European Union (Program ERB4001GT930714).
\vfill\eject
\noindent
{\bf REFERENCES}

{
\advance\leftskip by 4em
\parindent = -4em

[1] A. Maci\'as, O. Obreg\'on and M.P. Ryan, Class.~Quantum Grav.~{\bf 4},
1477 (1987).

[2] P.D. D'Eath and D.I. Hughes, Phys.~Lett.~{\bf 214}B, 498 (1988).

[3] R. Graham, Phys.~Rev.~Lett.~{\bf 67}, 1381 (1991).

[4] R. Graham, Phys.~Lett.~{\bf 277}B, 393 (1992).

[5] R. Graham and J. Bene, Phys.~Lett.~{\bf 302}B, 183 (1993).

[6] P.D. D'Eath and D.I. Hughes, Nucl.~Phys.~B {\bf 378}, 381 (1992).

[7] L.J. Alty, P.D. D'Eath and H.F. Dowker, Phys.~Rev.~D {\bf 46}, 4402
(1992).

[8] P.D. D'Eath, S.W. Hawking and O. Obreg\'on, Phys.~Lett.~{\bf 300}B, 44
(1993).

[9] P.D. D'Eath, Phys.~Rev.~D {\bf 48}, 713 (1993).

[10] M. Asano, M. Tanimoto and N. Yoshino, Phys.~Lett.~{\bf 314}B, 303 (1993).

[11] P.D. D'Eath, Phys.~Lett.~B, in press.

[12] C. Teitelboim, Phys.~Rev.~Lett.~{\bf 38}, 1106 (1977).

[13] P.D. D'Eath, Phys.~Rev.~D {\bf 29}, 2199 (1984).

[14] R. Graham and H. Luckock, University of Essen preprint.

[15] S.W. Hawking and D.N. Page, Phys.~Rev.~D {\bf 42}, 2655 (1990).

[16] J.B. Hartle and S.W. Hawking, Phys.~Rev.~D {\bf 28}, 2960 (1983).

[17] P.D. D'Eath, to appear in Phys.~Lett.~B.

[18] P.K. Townsend, Phys, Rev.~D {\bf 15}, 2802 (1977).

[19] M.P. Ryan and L.L. Shepley, {\it Homogeneous Relativistic Cosmologies}
(Princeton University Press, Princeton, 1975).

[20] J. Wess and J. Bagger, {\it Supersymmetry and Supergravity} (Princeton
University Press, Princeton, 1992).

[21] T. Sano and J. Shiraishi, University of Tokyo preprint UT-622; T. Sano,
University of Tokyo preprint UT-621.

[22] J.B. Hartle, in {\it High Energy Physics 1985}, ed. M.J. Bowick and F.
G\"ursey (World Scientific, Singapore, 1986).

[23] J.J. Halliwell and S.W. Hawking, Phys.~Rev.~D {\bf 31}, 1777 (1985).

[24] D.I. Hughes, Ph.D.~thesis, University of Cambridge (1990), unpublished.

[25] P.D. D'Eath and J.J. Halliwell, Phys.~Rev.~D {\bf 35}, 1100 (1987).

\ \ \ }
\bye